\documentclass[fleqn,12pt,twoside]{article}
\usepackage{espcrc1}

\usepackage{epsfig}
\usepackage[figuresright]{rotating}

\newcommand{\AmS}{{\protect\the\textfont2
  A\kern-.1667em\lower.5ex\hbox{M}\kern-.125emS}}
\def\be{\begin{equation}}
\def\ee{\end{equation}}
\def\bea{\begin{eqnarray}}
\def\eea{\end{eqnarray}}
\def\beq{\begin{equation}}
\def\eeq{\end{equation}}
\begin{document}

%\draft

%\null
%\begin{flushright}
%\end{flushright}
\title{Possible Origin of RHIC $R_{out}/R_{sid}$ HBT Results}
\author{Sandra S. Padula\address[MCSD]{Instituto de F\'{\i}sica Te\'orica, 
Universidade Estadual Paulista, S\~ao Paulo - SP, Brazil.}%
        \thanks{Theoretical Physics, FERMILAB, P.O. Box 500-MS106, Batavia, 
IL 60510, USA.}\thanks{Nuclear Theory Group, Brookhaven National 
Laboratory, Upton, NY 11973, USA.}}
\maketitle
\vskip-9cm\hfill
\begin{flushright}
%\centerline
BNL-69354\\
FERMILAB-Conf-02/227-T\\
IFT-P.070/02\\
\end{flushright}
\vskip 6.7cm
\begin{abstract}
The effects of opacity of the nuclei together with 
a blackbody type of emission along the system history 
are considered as a means to explain the ratio 
$R_{out}/R_{sid}$ observed by STAR 
and PHENIX collaborations at RHIC. Within our model, no flow is required 
to explain the data trend of this ratio for large surface emissivities. 
\end{abstract}
%\pacs{25.75.Gz,25.75.-q,74.80.Fp}

\vskip0.83cm

	The unexpected results presented by STAR\cite{star} in the 
previous Quark Matter meeting on  $\pi^\pm \pi^\pm $ HBT, 
later confirmed by PHENIX\cite{phenix}, regarding 
the decrease of the ratio $R_{out}/R_{sid}$ for increasing $K_T$, 
has been challenging explanations since that time. Hydrodynamic models 
and microscopic-based simulations usually predicted the opposite 
behavior for increasing pair momentum. 

	Motivated by this challenge we 
%Larry McLerran and I 
proposed a simple model 
(see \cite{mclpad} for details) to try and understand the unforseen 
decrease of the {\sl outwards} radius (along the direction of the average 
transverse momentum of the pair of pions, $K_T$) relative to the 
{\sl sidewards} one (i.e., orthogonal to $K_T$). Two were the main ingredients 
of this attempt. The first was to consider the particle emission 
%as from a blackbody,
of a blackbody type, radiating from the external surface of the system 
during its entire evolution. The second was to treat this system as 
an opaque source. The other basic ingredients were not unusual. The system 
produced in a heavy ion collision is supposed to be formed above the critical 
temperature, $T_c$, at the time $\tau_0$, in a Quark-Gluon Plasma (QGP) phase.
Its temperature gradually decreases while it expands and   
for the sake of simplicity, the expansion is considered to be only along 
the longitudinal direction. After this initial stage, lasting about 
($\tau_c - \tau_0$), where $\tau_c$ correspond to the on-set of the 
phase-transition at the temperature  $T_c$, 
the mixed phase begins, during which the temperature remains 
constant with time. The mixed phase continues for a longer 
period, ending after an elapsed interval ($\tau_h - \tau_c$). 
Then, the system converted into a gas of pions (no resonances are 
included) expands further, until 
the decoupling temperature, $T_f$, is reached. 
At this point, the system is quite dilute, since most  
of the particles have already been evaporated from its surface. Thus, 
at the time  $\tau_f$, the system is supposed to decouple in 
an instantaneous volumetric emission. No complex 
mechanism for the hadronization of quarks and gluons is considered 
in detail at this point, although hadronization must take place. 
In other word, in first approximation, the evaporation 
of ``gluons'' and ``quarks'' (as hadronized pions) from the external 
surface of the system is considered in the same way as emission of pions, 
except for the number of degrees of freedom.

The Bjorken hydrodynamical model\cite{Bj} is assumed to describe 
the system during its entire evolution, i.e., since it is formed and 
until it breaks up. This is supplemented 
with a blackbody type of radiation from the surface 
of the matter from its formation. The emitting source is supposed 
to be opaque, in a generalized version of opacity proposed by 
Heiselberg and Vischer\cite{henn} 
at CERN energies, later followed by Heinz and Tom\'a\u sik  
\cite{tomheiz1}.

To compute the emitted spectrum and the two particle distribution function, 
the Covariant Current Ensemble formalism, \cite{ccef_pgg}, is adopted. 
In this formalism, the two particle correlation function can be written as
$
	C(k_1,k_2) = {{P_2(k_1,k_2) } \over {P_1(k_1)P_1(k_2)}} 
= 1 + {{\mid G(k_1,k_2)} \mid^2 \over {G(k_1,k_1) G(k_2,k_2)}}
$, 
where $P_1(k_i)$  and $P_2(k_1,k_2)$ are, respectively, the single particle
distribution and the probability for simultaneous observation of 
two particles with momenta $k_1$ and $k_2$. The average and the relative 
momentum of the pair 
%is 
are defined as $ {\bf K} = ({\bf k_1} + {\bf k_2})/2$ and
%, and the relative momentum, as  
${\bf q} = {\bf k_1} - {\bf k_2}$. 
%, respectively. 

The emitted energy as well as the total entropy associated to each 
stage of the system evolution can be estimated. Just for a 
brief illustration, I write down the emitted energy as a 
function of time in the initial stage by 
considering the emission by an expanding cylinder 
of transverse radius $R_T$ and length $h$, in the 
time interval $\tau$ and $\tau+d\tau$, as  
%\beq
$ 
dE_{in} = - \kappa \sigma T^4 2 \pi R_T h d\tau - 
\frac{4}{3}  \sigma T^4 \pi R_T^2 dh
$, 
%\; \; , \label{denergy}\eeq
where the first term comes from the blackbody type of energy  
radiated from the external surface of the cylinder, and the second term 
results from the mechanical work due to its expansion. The $\kappa$
factor was introduced to take into account that the system has some opacity 
to surface emission.  The constant $\sigma$ is the Stefan-Boltzmann constant
and is proportional to the number of degrees  of freedom in the system. 
	By integrating this expression the energy density can be obtained as  
%i.e., $\epsilon = E/V$, as  
$\epsilon_{in} = \epsilon_0 (\frac{\tau_0}{\tau})^\frac{4}{3} 
e^{-\frac{2 \kappa}{R_T}(\tau-\tau_0)}
$.
From this expression it can be seen that the multiplicative 
factor, $e^{-\frac{2 \kappa}{R_T}(\tau-\tau_0)}$, appears 
in addition to that coming 
from the Bjorken picture. The variation of the temperature in the initial 
stage, i.e., prior to the beginning of the phase transition, follows 
immediately as
$
T(\tau) = T_0 (\frac{\tau_0}{\tau})^\frac{1}{3} 
e^{-\frac{\kappa}{2 R_T}(\tau-\tau_0)}
$. 
The instant corresponding to the beginning of the mixed 
phase, $\tau_c$, when $T_c=175$ MeV is reached, that one 
corresponding to its end, at $\tau_h$, as well as 
and the decoupling time, $\tau_f$, at $T_c=150$ MeV, are given by 
$
\tau_c \; e^{\frac{3 \kappa}{2 R_T} \tau_c} = 
\left( \frac{T_0}{T_c} \right)^3 \; \tau_0 \; 
e^{\frac{3 \kappa}{2 R_T} \tau_0} ; \; 
%\; \; , \label{tauc}\eeq
%\beq
\tau_h \; e^{\frac{2 \kappa}{R_T} \tau_h} = 
\left(\frac{g_g+g_q}{g_\pi} \right) \tau_c \; 
e^{\frac{2 \kappa}{R_T} \tau_c}  ; \; 
%\; \; , \label{tauh}\eeq
%\beq
\tau_f \; e^{\frac{3 \kappa}{2 R_T} \tau_f} = 
\left( \frac{T_c}{T_f} \right)^3 \; \tau_h \; 
e^{\frac{3 \kappa}{2 R_T} \tau_h} 
$.

Finally, the initial values of the 
temperature, $T_0$ and the formation time, $\tau_0$, are related to 
the initial entropy, $S_0$ and to the input number of particles, 
${\cal N}$ (chosen to match the average experimental pion multiplicity 
per unit of rapidity at RHIC, ${\cal N} \sim 1000$), by 
%\beq
$
S_0 = \Gamma {\cal N} = \left[ (g_{\small g} + g_{\small q}) \times
(\frac{4}{3}) \frac{\pi^2}{30} T_0^3 \right] \pi R_T^2 \; \tau_0 \; 
; \; \tau_0 \sim \frac{1}{3 T_0} (GeV^{-1}) \sim \frac{0.197}{3 T_0} (fm)
$, 
%\; \; , \label{N}\eeq
where $\Gamma = 3.6$, as estimated by the entropy per 
particle ($S_\pi/N_\pi$) of a pion gas at freeze-out. 
Then, $T_0 \sim 411$ MeV and $\tau_0 \sim  0.160$ fm. 
The degeneracy factors, $g_g + g_q$, account for the gluon and 
quark (antiquark) degrees of freedom ($g_g + g_q =37$ for two quark 
flavors). In the case of pions, the 
degeneracy factor is $g_{\small \pi} = 3$.  
I illustrate in the table below the time variables for two different 
assumptions on the emissivity, $\kappa$. I also write the estimated 
fraction of the particles emitted from the surface during the 
period $\tau_0 \le \tau \le \tau_f$, ${\cal S}/{\cal N}_{tot}$,  
relative to the total number of produced particles, ${\cal N}_{tot}$, 
as well as the remnant portion at freeze-out, 
{\large ${\cal V}$}$/{\cal N}_{tot}$, then emitted from 
the entire volume. 

\vfill\eject
%\vskip0.5cm
\begin{table}[htb]
%\caption{proper-time parameters $\tau_0$, $\tau_c$, 
%$\tau_h$, $\tau_f$, as well as of the surface, ${\cal S}$ 
%(for $\tau_0 \le \tau \le \tau_f$), and 
%the volume,  {\large ${\cal V}$} (at $\tau_f$), 
%emitted fluxes, for two values of the emissivity, $\kappa$}
\label{table:1}
\newcommand{\m}{\hphantom{$-$}}
\newcommand{\cc}[1]{\multicolumn{1}{c}{#1}}
\renewcommand{\tabcolsep}{1.2pc} % enlarge column spacing
\renewcommand{\arraystretch}{1.2} % enlarge line spacing
\begin{tabular}{@{}lllllll}
\hline
~ ~ $\kappa$  & $\tau_0$  & $\tau_c$ & $\tau_h$ & $\tau_f$ & 
${\cal S}/{\cal N}_{tot}$ & {\large ${\cal V}$}$/{\cal N}_{tot}$ \\
 &  (fm/c)  &  (fm/c) &   (fm/c)    & (fm/c)  & 
 ($\tau_0 \le \tau \le \tau_f$) & (at $\tau_f$) \\ 
\hline
~ ~ 1     & 0.160 & 1.54 & 5.73 & 6.97 & 0.844 & 0.156 \\
~ ~ 0.5   & 0.160 & 1.75 & 8.37 & 10.5 & 0.758 & 0.242 \\
\hline
\end{tabular}\\[2pt]
%%The experimental values are given in ref. \cite{Eato75}.
\end{table}

\vskip-0.5cm

	In order to check how the spectra estimated within our model behave 
as compared to data (PHENIX minimum bias
\cite{phenix2}), I plot its predictions on the 
single-inclusive distribution in Fig. 1(a). The 
estimates and discussions presented here are restricted to the central 
rapidity region, i.e., 
$y_i = 0$ (which implies that $k_{i_L} = 0$, and, consequently, $K_L = 0$ 
and $q_L = 0$).  In Fig. 1(b) I show the results for the ratio 
%of the {\sl outwards} by {\sl sidewards} radii, 
$R_{out}/R_{sid}$ vs. $K_T$, together with the preliminary STAR\cite{star}
(filled triangles) and PHENIX\cite{phenix} (filled circles) data for both 
$\pi^+ \pi^+$ and $\pi^- \pi^-$ interferometry. 

%\noindent
\null
%\vskip-6cm
\begin{figure}[htbp]
% \vspace{2mm}
\vspace{-1.1cm}
%  \vspace{-2.1cm}
 \begin{minipage}{0.49\textwidth}
 \includegraphics[angle=-90,scale=0.41]{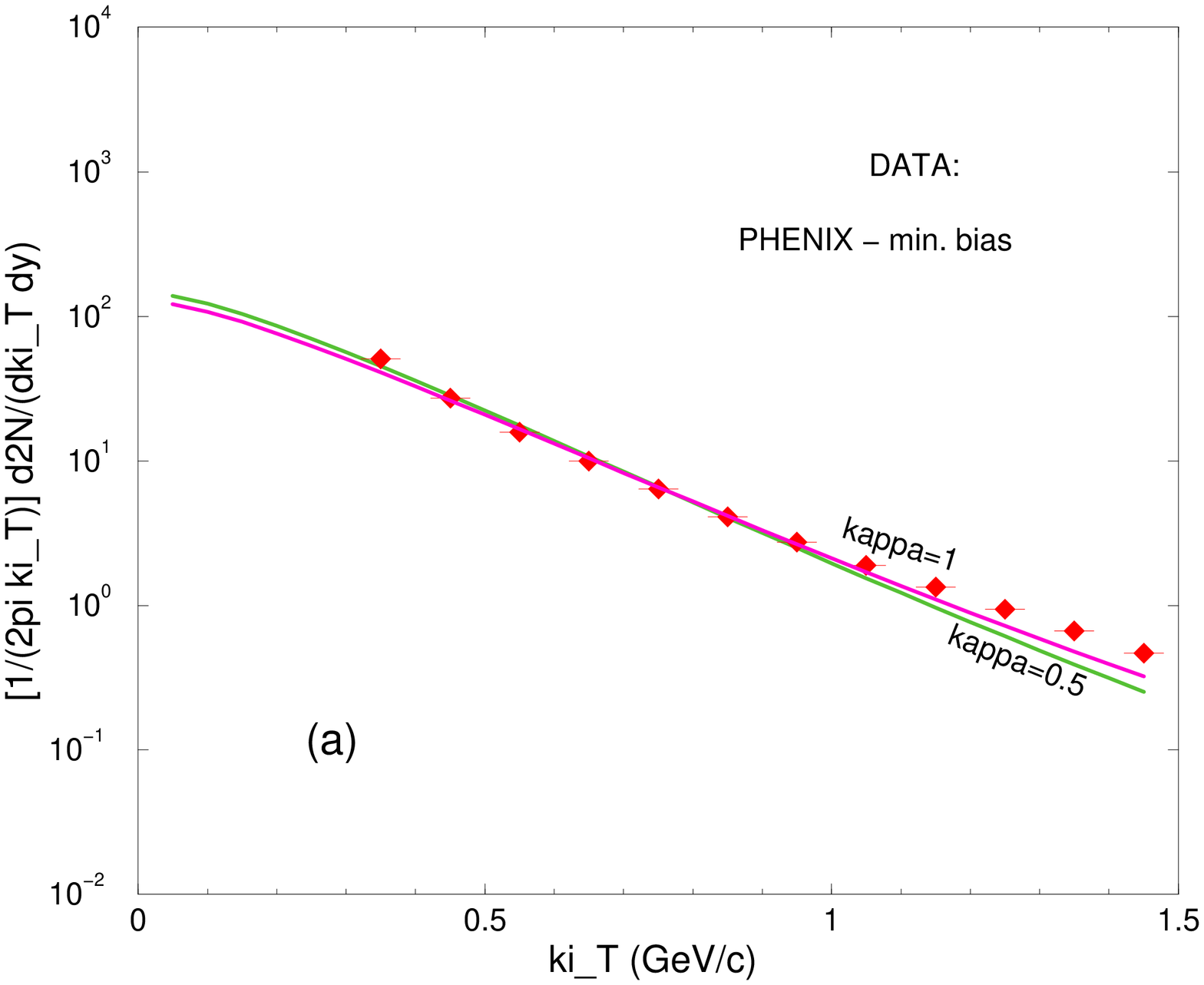} 
 \end{minipage}
 \hfill
 \begin{minipage}{0.49\textwidth}
  \includegraphics[angle=-90,scale=0.41]{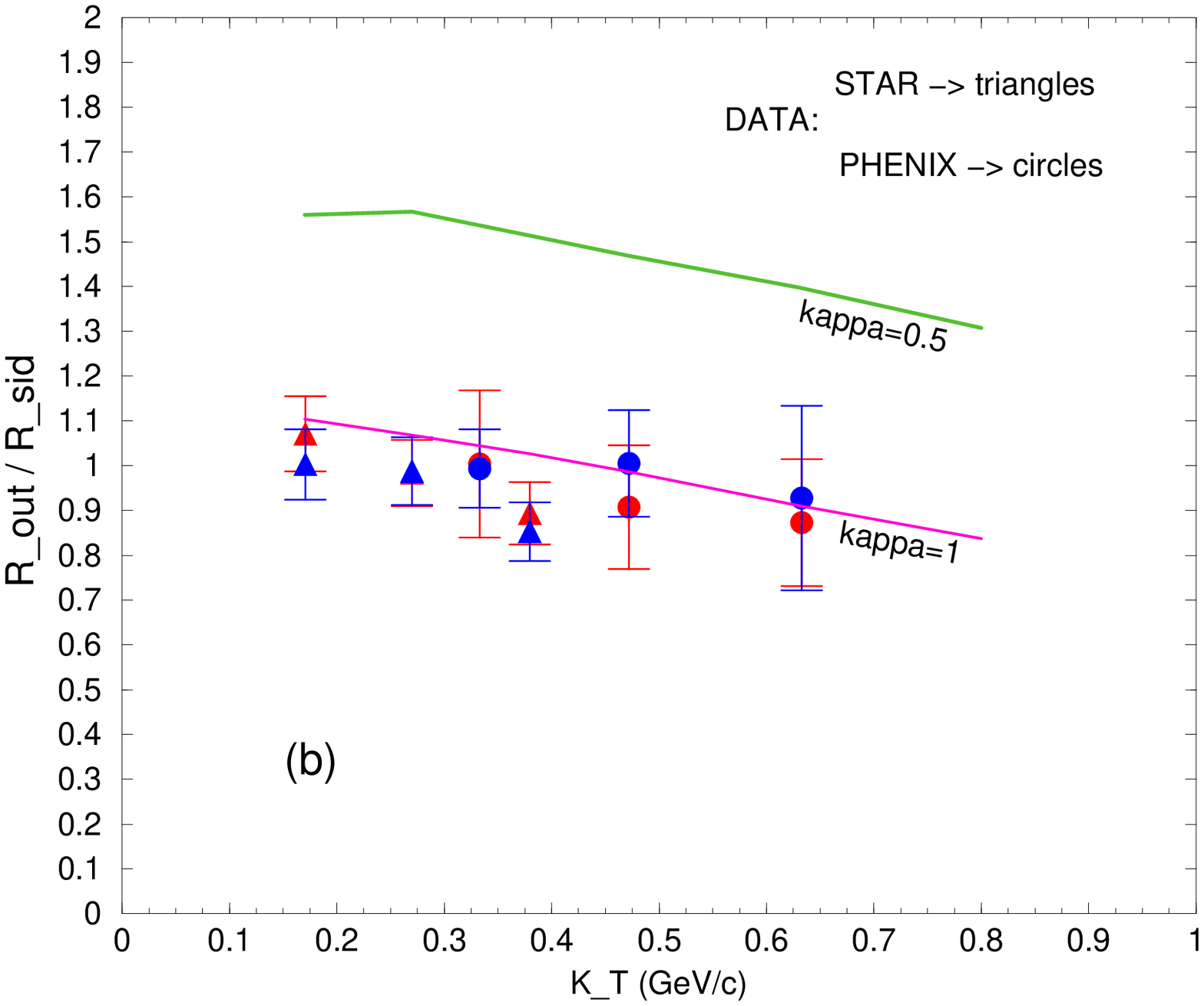} 
 \end{minipage}
 \vspace{-1cm}
  \caption{{\bf (a):} The prediction based on our model for  the transverse momentum 
distribution of emitted pions is shown. The points are from the minimum-bias 
data from PHENIX Collaboration. The curves correspond to 
emissivity $\kappa=0.5$ and to $\kappa=1$, without the inclusion of any 
transverse flow. Both cases describe data on spectrum well 
in the low pion momentum region, up to about  $k_{i_T}\approx 1$ GeV/c.  
   {\bf (b):} The results for the ratio $R_{out}/R_{sid}$ of the 
{\sl outwards} by the {\sl sidewards} radii are shown within our model. 
The ratio corresponding to full emissivity ($\kappa=1$) agrees very well 
with data within the experimental 
error bars (shown in the plot), whereas the $50\%$ emissivity case is 
completely excluded by data. 
The parameters used are given in the text and in the table, corresponding 
also to $T_0 = 411$ MeV, $T_c = 175$ MeV, $T_f = 150$ MeV, 
and the transverse radius, $R_T \approx 7$ fm/c.}
\label{Coh}
\end{figure}

\vskip-0.9cm
As seen from the plots in Figure 1, our results were extremely 
successful in 
describing both sets of data for $\kappa=1$, but the curve corresponding 
to $\kappa=0.5$ is away above the data limits, suggesting that there should 
be high emissivity along the system history in order to explain the 
data trend.  
The model also describes the typical source radii reasonably
well, but not the $K_T$ dependence of these radii (see \cite{mclpad} for 
details). This suggests that the time variation of the emitting radius and 
the introduction of transverse flow may play a significant role, 
\cite{heinzkolb}-\cite{teaney}. 
If there is a time variation of the various radii, this will be
correlated with the typical momentum scale of emitted particles, since the
earlier is the time, the hotter are the particles.  Our model is sensitive
to such variation since emission from the hot surface is allowed at early times.
Also, a proper treatment of the decoupling is not included in our 
computations, which would affect the results, although it might 
also suggest modification in the treatment of decoupling, 
\cite{teaney},\cite{bass}.

The principal reason why such a small ratio of $R_{out}/R_{side}$ vs. $K_T$ 
is obtained within our model is probably due to a combination of two effects.
The first is that the surface is opaque, and whatever is emitted from the 
surface will have a small value of this radius.  The second effect 
is that black body radiation by partons is allowed when the surface
is very hot.  This allows a much larger contribution from surface 
emission than is typical of what happens in hydrodynamical simulations, where
particles are emitted by Cooper-Frye decoupling from a surface at very low
temperature.  In fact about $80\%$ of the emission comes from the surface 
in our model.  The fact that so many particle are emitted from the 
surface at early times also means that the longitudinal decoupling time
in this computation is significantly shorter than would 
be the case for hydrodynamic simulations. I should add that many of the features
of the model proposed and discussed here are embodied in the hydrodynamic 
computations of Heinz and Kolb\cite{upkolb}.  The main difference lies in 
the treatment of an essential ingredient, the emissivity of the surface.

%\bigskip
%\noindent
%Acknowledgments

\vskip1cm
I am deeply grateful to Larry McLerran and 
the Nuclear Theory Group at BNL, as well as to Keith Ellis and 
the Theoretical Physics Department 
at Fermilab, for their kind hospitality 
during the elaboration of this work. 
This research was partially supported by CNPq (Proc. N. 200410/82-2). 
This manuscript has been authored under Contracts No. DE-AC02-98CH10886 
and No. DE-AC02-76CH0300 with the U.S. Department of Energy.

\end{document}